\documentstyle[preprint,prb,aps,epsf]{revtex}   
\begin{document}
\draft
\title{Magnetic sublayers effect on the exchange coupling
oscillations {\it vs.} cap-layer thickness}

\author{M.~Zwierzycki and S.~Krompiewski{\footnote[2]{corresponding
author, fax: (48-61) 8684524, e-mail: stefan@ifmpan.poznan.pl}}} 
 
\address{ Institute of Molecular Physics,
P.A.N., Smoluchowskiego 17, 60-179 Pozna\'n, Poland}

\maketitle
\centerline {(Received .................. 1997)}

\begin{abstract}
We have found that some periods of interlayer exchange coupling (IEC)
oscillations as a function of cap-layer (CL) thickness may be suppressed if
the in-plane 
extremal spanning vectors of the cap- and ferromagnet-materials Fermi surfaces
do not coincide. The suppression of the IEC oscillations {\em vs.} the
CL thickness holds also if the magnetic slab thickness tends to infinity.
On the one hand, we have shown by means of very simple arguments
that apart
from the well-known selection rules concerning the spacer- and  cap-layers,
another one related with the magnetic sublayers has to be fulfilled in
order that
the interlayer coupling oscillations $vs.$ CL thickness could
survive. On the other hand, the distribution of 
induced magnetic moments across the non-magnetic cap- and spacer-sublayers
have been computed and shown to reveal the underlying
periodicity of the materials they are made of (i.e. related to their
bulk Fermi surfaces) independently of whether or not the selection
rules are fulfilled. This means that the IEC oscillations are of
global nature and depend on all the sublayers the system consists of.
 
\end{abstract}
\vspace*{4mm}

\section{Introduction}
Magnetic multilayers have been intensively studied for over a decade now
\cite{Grunberg,Parkin,Chappert}. The reasons for it, apart from challenging
cognitive aspects, are (already partially realized) practical applications
of superlattices as magneto-resistive sensors, angular velocity
meters, recording
heads and magnetic memory elements. The phenomenon most of these applications
is based on is the well known giant magneto-resistance (GMR) coming
from a strong
electron-spin dependence of resistivity in magnetic systems.
To optimize devices of that sort, it is necessary to test the effect of all
of the ingredients of the system in question (including kind of
materials they are made from and thicknesses of particular sublayers),
either directly on GMR or indirectly on the interlayer exchange
coupling (IEC). Obviously, the effect of a spacer on IEC was
established first \cite{Chappert,Edwards}, the next in 
turn was that due to magnetic sublayers
\cite{Barnas,SK,SKFSUK,Bruno_M,Bloemen,Okuno_M} and finally the 
cap-layer (CL) effect has been studied quite recently
\cite{Vries,Okuno_Cap,Barnas_Cap,Bruno_Cap,Kudrnovsky}.

Before we present our original results let us briefly recall
what are the most important facts
concerning the CL's~:
i)~the IEC oscillates as a function of CL thickness with a period determined
by extremal  $\vec{k}$ spanning vectors of the CL Fermi surface,
ii)~a bias of the oscillations (their asymptotic value) depends on  
spacer thickness \cite{Vries,Bruno_Cap,Kudrnovsky},
iii)~the IEC oscillations are strongly suppressed if stationary in-plane
spanning vectors of the CL Fermi surface do not
coincide with 
their counterparts of the spacer Fermi surface\cite{Bruno_Cap,Kudrnovsky},
iv)~the direct-  and inverse-photoemission\cite{Ortega,Ortega1} 
on various combinations of
overlayers deposited on different films shows a periodic distribution of the
so-called quantum well states (QWS) with periods determined by extremal
spanning vectors of the overlayer Fermi surface. We shall refer to the latter
only indirectly, by exploiting the fact that the QWS lead
to some spin-polarization of non-magnetic cap-layers.

The aim of the present paper is to emphasize the relevance of magnetic
sublayers to IEC oscillations as a function of cap-layer (CL)
thickness. Besides, we shall comment on induced magnetic moments in
the non-magnetic sublayers, which may be viewed as a manifestation
of the quantum well states \cite{Ortega,Ortega1,Cu_polar,Segovia}.
  
\section{Method}

Our earlier papers \cite{SK,SKUK,SKMZUK} based on the single-band
tight-binding model have proved that the model we use gives a reasonable
qualitative
description of basic physical mechanisms responsible for oscillatory
phenomena in magnetic trilayers.

Our Hamiltonian, described in detail in Ref.~\onlinecite{SKMZUK},
consists of the 
nearest-neighbor hopping and spin-dependent on-site
potential terms.
The  systems under consideration now are trilayers capped with an overlayer,
of the type
$n_{ovr} O/n_f F/n_s S/n_f F$, where $n_{ovr}$, $n_f$ and $n_s$ stand for the
numbers of cap- (O), ferromagnet- (F) and spacer- (S)
monolayers in the perpendicular $z$-direction. Hereafter the subscripts
and superscripts $ovr$ and $s$ will always refer to the cap-
and spacer-layers,
whereas the spin-dependent parameters referring to ferromagnetic sublayers
will be indexed by $\sigma = \uparrow$ or $\downarrow $. For simplicity,
we restrict ourselves to a simple cubic structure and regard the lattice
constant
and the hopping integral as the length- and energy-units, respectively.

The interlayer exchange coupling has been calculated from the difference in
thermodynamic potentials exactly as in Ref.~\onlinecite{SKMZUK}, moreover
the magnetic moments
(including the induced ones), $m$, have been expressed in terms of 
the eigen-functions $u$ of the Hamiltonian as 
$ m_i = n_{i\uparrow} - n_{i\downarrow}$,   
with
$n_{i \sigma} = \sum \limits_{E}|u _{i,\sigma} (E)|^2$,
where the summation runs over occupied states.	

\section{Asymptotic limits}
In this section we present some analytic formulae which will be useful
for interpretation of rigorous numerical resluts of the next section.
As has been shown in Ref.~\onlinecite{Mathon}, the IEC can be
Fourier-transformed with
respect to $n_{s}$ and $n_{\sigma}$. 
That procedure can be quite straightforwardly generalized to include
CL thickness as well. The resulting asymptotic (within the
stationary phase approximation) expression consists of the terms of
the form $A_{pqrn}(\vec{k}_{\|},E_F)exp\left(2i(pk^s_zn_s+(qk^{\uparrow}_z+
rk^{\downarrow}_z)n_f+nk^{ovr}_zn_{ovr})\right)$ summed over all the
in-plane wave vectors for which the exponential is stationary.
The $A$ coefficients are defined analogously as in
Ref.~\onlinecite{Mathon}.  Their exact numerical values are not
important for qualitative considerations, one notes only that all the
amplitudes of oscillations vanish asymptotically with the given
sublayer thickness going to infinity\cite{Mathon}. There exist, however,
some additional restrictions imposed by the asymptotic behavior of the IEC.
In particular, a
direct generalization of the results of Ref.~\onlinecite{Mathon} to the
present case, with the cap-layer, gives:
$A_{0qrn}=0$ (no coupling for $n_s \rightarrow \infty$).
Another limit to be taken is $n_f \rightarrow \infty$, when, in view of
the above mentioned asymptotic behavior, all the terms tend to
zero except for $A_{p000}$ and $A_{p00n}$. Since the oscillations 
{\em vs.} spacer thickness survive in this limit in contrast to the ones 
{\em vs.} CL thickness which decay (see below), we conclude that
$A_{p000}\ne 0$ and $A_{p00n}=0$.
 
Finally, taking into account the above mentioned restrictions
and keeping for simplicity only the lowest order harmonics, 
we arrive at the following formula~:
\begin{eqnarray}
\label{eq:J1}
J&=&\sum_{\alpha} A_{1000}e^{2ik^{s}_{z}n_{s}}+\sum_{\alpha_1}
A_{1100}e^{2i(k^{s}_{z}n_{s}+k^{\uparrow }n_{f})}+\sum_{\alpha_2}
A_{1010}e^{2i(k^{s}_{z}n_{s}+k_{z}^{\downarrow }n_{f})} \nonumber \\
~&~&+\sum_{\alpha_3} A_{1101}e^{2i(k^{s}_{z}{n}_{s}+k^{\uparrow}_{z}n_{f}+
k_{z}^{ovr}n_{ovr})}+
\sum_{\alpha_4} A_{1011}e^{2i(k_{z}^{s}n_{s}+k_{z}^{\downarrow
}n_{f}+k_{z}^{ovr}n_{ovr})}+\ldots,
\end{eqnarray}
where the $\alpha$-s are the sets of in-plane wave vectors for which
the relevant
exponentials are stationary. 
For the case of the CL thickness dependence this allows us to formulate
the following new selection rule, which in its general form (for
$n_s$ and $n_f$ large and fixed and $n_{ovr}$ large and varying) reads~:
\begin{equation}
\nabla k^{ovr}_z=0,\;\; p\nabla k^s_z n_s + (q \nabla k^{\uparrow}_z
+r \nabla k^{\downarrow}_z)n_f=0,
\label{eq:2}
\end{equation}
with nonvanishing $p$ and either $q$ or $r$ ($\nabla$ is the two-dimensional
gradient in the $k_x$,$k_y$ space).
This means that out of all the stationary vectors of the cap material FS
only those which simultaneously satisfy the above mentioned conditions for
the in-plane gradients give rise to the oscillations with CL thickness. 
Eq.~(\ref{eq:2}) is the main result of the present paper.
This condition becomes even simpler in the particular case of the 
single--band simple cubic model considered hereafter, when the second 
part of Eq.~(\ref{eq:2})
separates and all the individual in-plane gradients must vanish
(c.f. Ref.~\onlinecite{Mathon}).

The origin of the new selection rule becomes clear if we
qualitatively interpret Eq.~(\ref{eq:J1}) in terms of the quantum
interference model\cite{Bruno_I}. The first term corresponds to the
states reflected once at each of the spacer-ferromagnet interfaces, the
second and third ones to the states penetrating one of the magnetic layers
and reflected  back at the cap-ferromagnet interface while the last two
terms describe
states reaching the outer boundary of the cap-layer (``vacuum''). It
is quite clear therefore that
the $n_f$-dependent phase  factor must be also taken into account
while performing the stationary phase approximation.

It is evident from formula (\ref{eq:J1}) that the  bias of oscillations
with CL thickness depends not only on the spacer- and magnetic- layer
thicknesses but on the
on-site $V^{ovr}$ potential as well. The latter
observation results
from the fact that the $A$ coefficients in the second and third terms
of Eq.~(\ref{eq:J1})
depend on the value of the reflection coefficient at the cap-ferromagnet
interface which in turn depends on the cap material electronic structure.

The stationary spanning vectors, for a
sublayer characterized by the potential $V$, can be
determined in a very simple way, by minimizing with respect to
$\vec{k}_{\|}$ the following Fermi surface equation for the s.c. lattice~:
\begin{equation}\label{eqkz}
k_z(\vec{k}_{\|},E_F)=\arccos [ (V-E_F )/2 - \cos k_x -\cos k_y ]\,.
\label{kz1}
\end{equation}
Hence the in-plane extremal spanning vectors are:
$\vec{k}_{\|} = (0,0)$ for $-6 < E_F-V <-2$; $(\pm \pi ,0)$
$(0, \pm \pi )$ for $-2<E_F-V<2$; $(\pm \pi ,\pm \pi)$ for $2<E_F-V<6$
and
\begin{equation}\label{eqkz2}
k_z = \arccos [ (V-E_f )/2 - \alpha ] \,,
\label{kz2}
\end{equation}
with $\alpha=$ 2, 0 and -2, for the corresponding $\vec{k}_{\|}$,
respectively. Thus, the period of oscillations 
$vs.$ sublayer (with the potential $V$) thickness, is just $\Lambda=\pi/{k_z}$
(or $\pi/{(\pi - k_z)} $). 

\section{Numerical results}
\begin{figure}[h]
  \epsfxsize=16.5cm
  \epsfbox{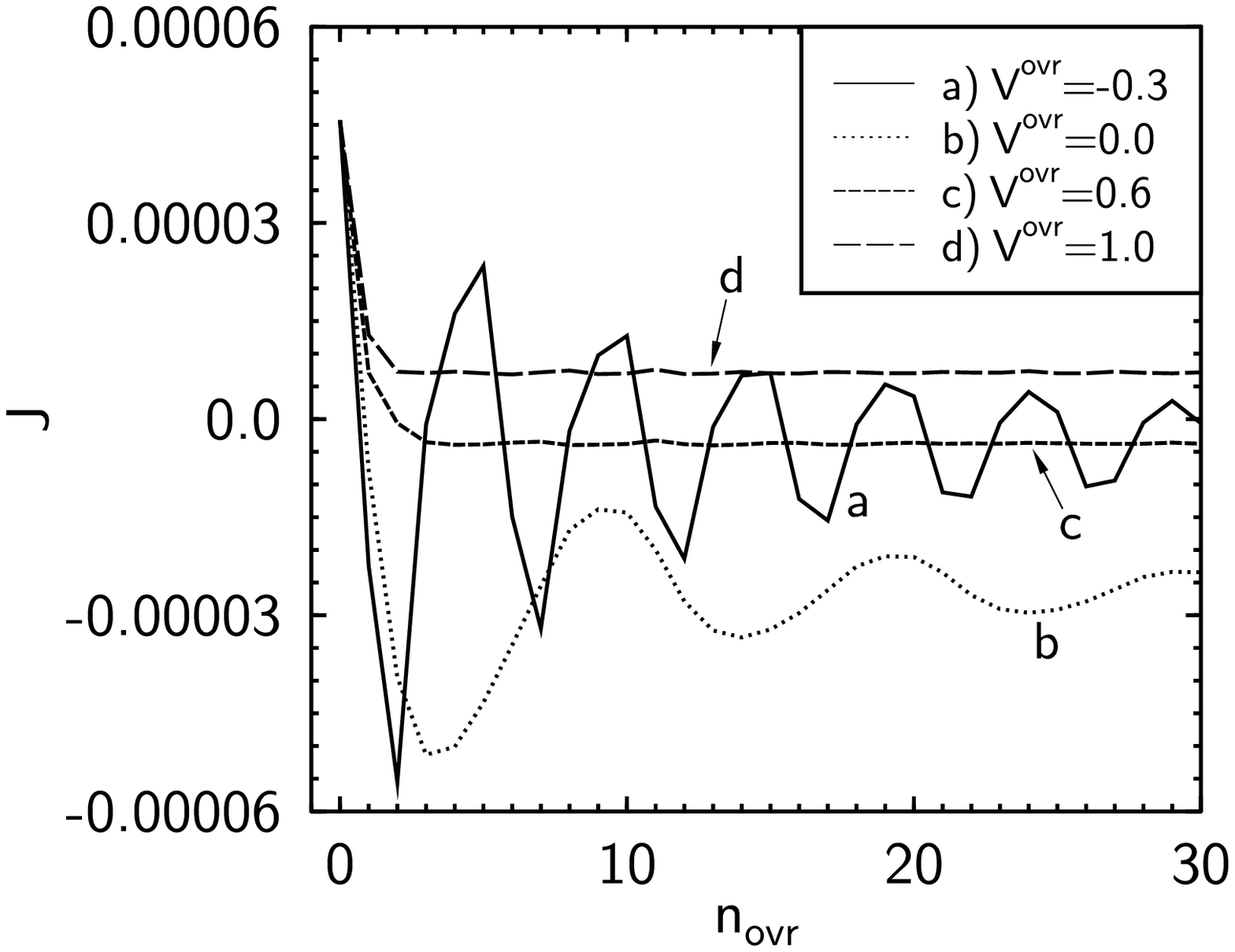}
\caption{The exchange coupling {\em vs.} cap layer thickness for
$n_s=5$, $n_f=3$, $E_F=2.1$, $V^{s}=V^{\downarrow}=0$ and
$V^{\uparrow}=-2.0$. Stationary in-plane spanning vectors of the
spacer and both
the ferromagnetic FS's are $k_{\|}=(\pm \pi,\pm \pi)$. For
the curves $a$ and $b$ the stationary in-plane vector of the cap FS is the same
in contrast with the $c$ and $d$ cases, for which 
$k_{\|}=(0,\pm \pi ),( \pm\pi,0)$ 
what results in suppressing the oscillations.}
\label{fig1}
\end{figure}

\begin{figure}[h]
  \epsfxsize=16.5cm
  \epsfbox{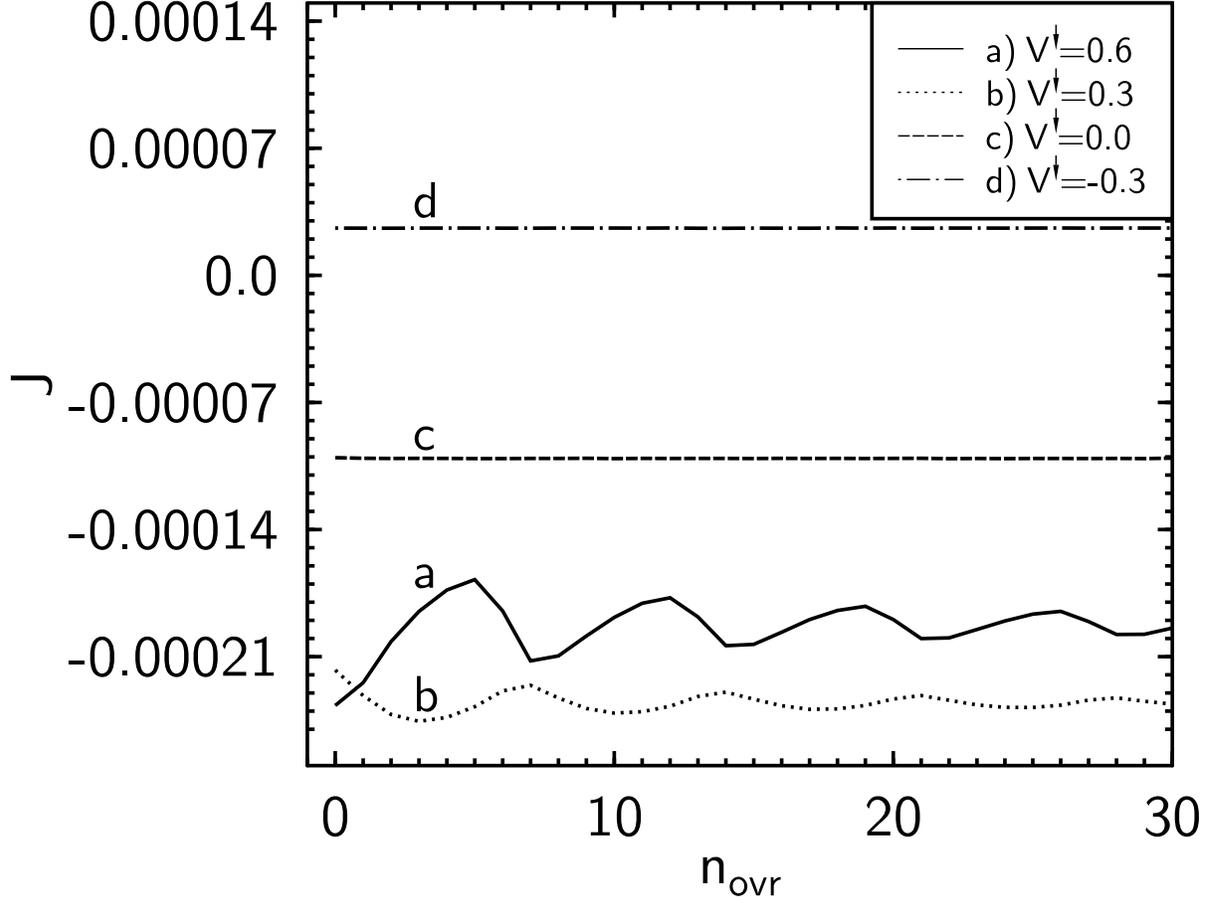}
\caption{IEC {\em vs.} CL thickness for $n_s=5$, $n_f=10$, $E_F=2.1$,
$V^{s}=V^{ovr}=0.3$ and $V^{\uparrow}=-2.0$. Stationary in-plane
spanning vectors for the
spacer and the cap-layer are $k_{\|}=(0,\pm \pi),\, (\pm\pi,0)$. 
For the first two curves ($a$ and $b$) the
minority spin FS stationary points coincide with those of the spacer and
the overlayer. In case of the curves $c$ and $d$ both the majority and
minority spin Fermi surfaces have the $k_{\|}=(\pm \pi,\pm \pi)$
spanning vector and consequently the IEC oscillations are
suppressed.}
\label{fig2} 
\end{figure}

\begin{figure}[h]
  \epsfxsize=16.5cm
  \epsfbox{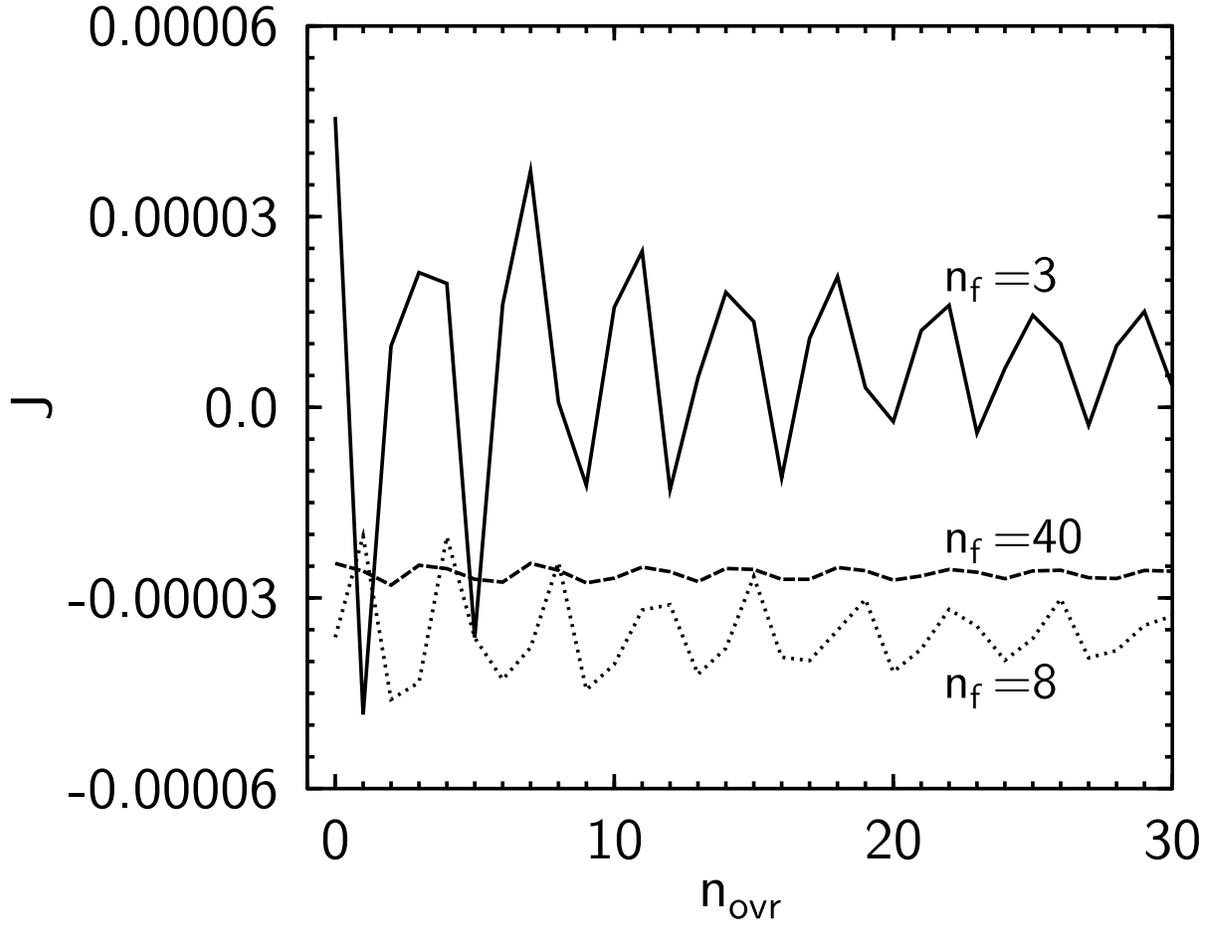}
\caption{The effect of magnetic sublayer thickness on the IEC
oscillations as a function of cap-layer thickness (parameters as in
Fig.~1 except for $V^{ovr}=-0.6$).}
\label{figa} 
\end{figure}

\begin{figure}[h]
  \epsfxsize=16.5cm
  \epsfbox{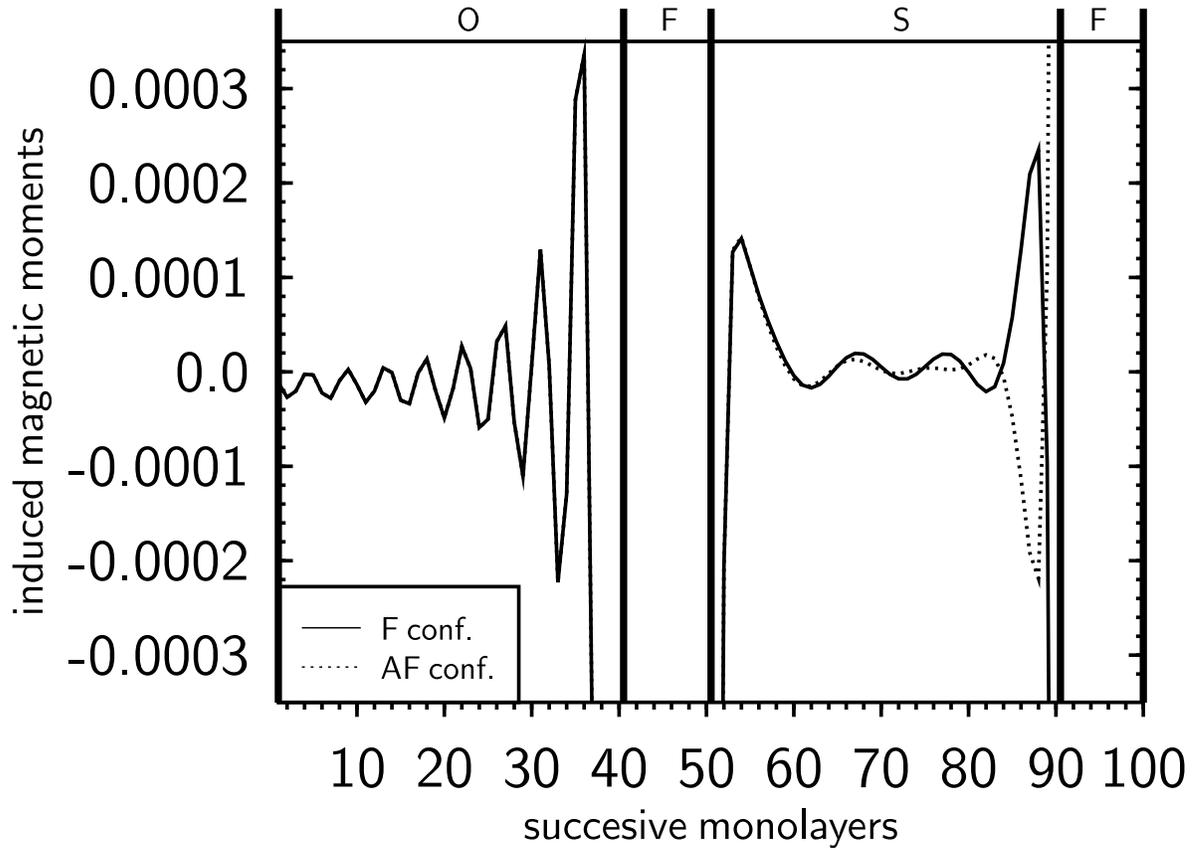}
\caption{Induced magnetic moments (with $\mu_B=1$) for $n_s=n_{ovr}=40$ and
$n_f=10$ for parallel (full line) and antiparallel (dashed line)
configurations. The other parameters as in Fig.~1c. Thick vertical
lines mark the interfaces. 
}
\label{fig4}
\end{figure}

We shall now present our exact numerical results (see
Ref.~\onlinecite{SKMZUK} for
details of the method) and show how they can be interpreted in terms of
the analytical formulae from the preceding section.
Fig.~\ref{fig1}. confirms the well-known fact that the IEC
oscillations {\it vs.} CL thickness have got
a period determined by kind of material the cap is made of and get suppressed
if there is a mismatch in corresponding in-plane spanning vectors of
the CL and the spacer.
The suppression takes place in the cases $c$ and $d$
where ${\vec{k}_{\|}}^s={\vec{k}_{\|}}^{\sigma}=({\pm \pi,\pm \pi})$,
opposite to ${\vec{k}_{\|}}^{over}=({0,\pm \pi})$,$(\pm \pi,0)$. The
dependence of the bias values on $V^{ovr}$ is also clearly visible.
The new effect is presented in Fig.~\ref{fig2}, which shows that the
suppression may be
due  to the misfit in the ${\vec{k}_{\|}}$'s corresponding to the overlayer and
magnetic sublayers, respectively (curves $c$ and $d$), whereas in case
of the curves 
$a$ and $b$ the periodicity is quite pronounced owing to the matching of
the above mentioned spanning vectors. It can be also readily seen from
Fig.~\ref{fig2} that the phases of oscillations as well as the bias-values
depend on the potentials of the ferromagnetic layers (exchange
splitting).
It is noteworthy that Figs.~\ref{fig1} and \ref{fig2} show that the
selection rule works quite
well, even when the relevant layer thicknesses are rather small: $n_s=5$ and
$n_f=10$, respectively. This confirms our previous 
observation\cite{SKMZUK} that 
relatively small systems in the z-direction may reveal the asymptotic
behavior. A detailed inspection of curves c and d suggests that the selection
rule is slightly more rigorously enforced in Fig.~\ref{fig2} (due to $n_f=10$) than
in Fig.~\ref{fig1} (due to $n_s=5$), but the effect is tiny indeed and hardly visible.
Incidentally, all the periods of oscillations obtained by the
numerical computations and visualized in
Figs.~\ref{fig1}--\ref{fig4} can be pretty well
reproduced in terms
of the asymptotic Eqs.~(\ref{kz1}) and~(\ref{kz2}): e.g. for $E_F=2.1$ and
$V=-0.6,\,-0.3,\,0.0,\,0.3$ and $0.6$, we get $\Lambda=3.6,\,4.9,\,9.9,\,6.9$
and $4.3$~ML respectively.

Another rather obvious but noteworthy effect consists 
in vanishing of the IEC
oscillations {\em vs.} CL thickness when magnetic sublayer thickness
gets bigger and bigger. This is shown in Fig.~\ref{figa}, and has not been
discussed either, to our knowledge so far, although such a trend could be
predicted on the basis of analytical formulae of
Ref.~\onlinecite{Bruno_Cap}. In fact this finding means that in order to
avoid undesirable effects of cap-layers (which may be of different
thickness in an experiment) on the IEC oscillations one should work
with thick magnetic sublayers. It is also noteworthy that the
oscillation bias
value depends on the magnetic layer thickness, as could be predicted from
Eq.~\ref{eq:J1}.

Finally, in connection with the quantum well states
concept\cite{Ortega,Ortega1,Cu_polar,Segovia}, we have studied the
distribution of induced magnetic
moments in the CL (and in the spacer). A typical result is presented
in Fig.~\ref{fig4}.
The induced magnetic moments are measured in dimensionless units 
($\mu_B = 1$)
and are of the order of 0.1\% with respect to the magnetic layer
magnetization.
As expected, the period of the induced-moment distribution within the CL is
exactly that anticipated for the bulk CL material FS.
The effect of the other sublayers is minor, except that the magnitude
of the induced moments is also magnetic-slab dependent. 
This might seem, at a first glance, to be in conflict with the IEC behavior
which shows no oscillations for parameters of Fig.~\ref{fig4}
(cf. Fig.~\ref{fig1}c).
Yet the spin polarization in non-magnetic layers is related to just one system
with the fixed sublayer thicknesses and the given alignment of magnetic
sublayers, whereas
the IEC results from the total energy (thermodynamic potential)
balance between 
the two possible ferromagnetic layer alignments and has to do with the series
of samples with changing CL thicknesses. This observation implies that the
induced-magnetic moments in the non-magnetic cap-layer (as well as the QWS)
give in general the whole set of periods, out of which only those survive,
as far as the IEC is concerned, which fulfill the selection rules referring to
the entire system. In other words the IEC oscillations are the global
characteristic of the whole system, whereas the induced spin
polarization in the cap-layer is strictly of local nature.

The selection rules completed hereby by the extra condition
related with the extremal spanning vectors of the magnetic sublayers,
are quite general and apply to real systems, too. In particular they allow to
explain why in case of the Cu/Co/Cu/Co multilayer the short period of
oscillations with Cu cap-layer thickness
is absent\cite{Vries} in  spite of theoretical predictions
\cite{Bruno_Cap} and the photoemission
results concerning QWS \cite{Ortega,Ortega1,Segovia}. In fact
the explanation is simple and quite
analogous to that of Ref.~\onlinecite{Mathon} about  IEC oscillations as 
a function of
ferromagnetic layer thickness. Out of two in-plane extremal spanning vectors
of the Cu Fermi surface only the ``belly'' one (at $ \vec{k}_{\|}=0 $ )
coincides with the extrema of the majority and minority sheets of the
Co Fermi surface, giving rise to the long period of oscillations.
The ``neck'' spanning vector has no counterpart in the Co FS and this is why 
there are no short period oscillations.

In conclusion, we have shown that in order for the interlayer exchange
coupling oscillations {\em vs.} cap-layer thickness to exist, it is
necessary, that both the cap-layer- and magnetic-layer- Fermi surfaces
share the same extremal in-plane spanning vectors. If this new
``selection rule'' is not fulfilled the period anticipated from the
bulk cap-layer material will not occur in the exchange coupling,
although it will still be present in the induced moment distribution
across the cap layer. Another finding of this paper is that the IEC
 oscillations {\em vs.} CL - thickness vanish if magnetic sublayers
thickness tends to infinity.

This work has been carried out under the KBN grants
No.~2P03B~165~10 (MZ) and  2-P03B-099-11 (SK). We thank the
Pozna\'n Supercomputing and Networking Center for the computing time.

\newpage
\listoffigures
\newpage
\newpage
\newpage
\newpage
\end{document}